\documentclass[aps,prx,twocolumn,superscriptaddress,showpacs,showkeys]{revtex4-2}

\usepackage[utf8]{inputenc}
\usepackage{graphicx}
\usepackage{amsmath}
\usepackage{amssymb}
\usepackage{array}
\usepackage{multirow}
\usepackage{float}
\usepackage{color}
\usepackage[normalem]{ulem}
\usepackage[T1]{fontenc}
\usepackage{hyperref}
\hypersetup{breaklinks=true,hidelinks,colorlinks=true,urlcolor=blue,citecolor=blue,linkcolor=blue}

\begin{document}
\newcommand{\rixscd}{$[I_{\rm{CL}}-I_{\rm{CR}}]/\sum\limits_{\omega}[I_{\rm{CL}}+I_{\rm{CR}}]$}
\newcommand{\rixscdmean}{$I_{\rm{diff}}/\bar{I}_{\rm{sum}}$}
\newcommand{\rixsasmean}{$\bar{I}_{\rm{diff}}/\bar{I}_{\rm{sum}}$}

\newcommand{\rixsas}{$\sum\limits_{\omega}(I_{\rm{CL}}-I_{\rm{CR}})/
\sum\limits_{\omega}(I_{\rm{CL}}+I_{\rm{CR}})$}
\newcommand{\pss}{\sigma\sigma}
\newcommand{\psp}{\sigma\pi}
\newcommand{\pps}{\pi\sigma}
\newcommand{\ppp}{\pi\pi}

\title{Circular dichroism in resonant inelastic x-ray scattering from birefringence in CuO}

\author{Abhishek Nag}
\email{abhishek.nag@ph.iitr.ac.in}
\thanks{These authors contributed equally to this work.}
\affiliation{Center for Photon Science, Forschungsstrasse 111, Paul Scherrer Institute, 5232 Villigen-PSI, Switzerland}
\affiliation{Department of Physics, Indian Institute of Technology Roorkee, Uttarakhand 247667, India}
\author{G\'erard Sylvester Perren}
\thanks{These authors contributed equally to this work.}
\affiliation{Center for Photon Science, Forschungsstrasse 111, Paul Scherrer Institute, 5232 Villigen-PSI, Switzerland}
\author{Hiroki Ueda}
\affiliation{Center for Photon Science, Forschungsstrasse 111, Paul Scherrer Institute, 5232 Villigen-PSI, Switzerland}
\author{A. T. Boothroyd}
\affiliation{Department of Physics, University of Oxford, Clarendon Laboratory, Parks Road, Oxford OX1 3PU, UK.}
\author{D. Prabhakaran}
\affiliation{Department of Physics, University of Oxford, Clarendon Laboratory, Parks Road, Oxford OX1 3PU, UK.}
\author{M. Garc\'ia-Fern\'andez}
\affiliation{Diamond Light Source, Harwell Campus, Didcot OX11 0DE, United Kingdom}
\author{S. Agrestini}
\affiliation{Diamond Light Source, Harwell Campus, Didcot OX11 0DE, United Kingdom}
\author{Ke-Jin Zhou}
\affiliation{Diamond Light Source, Harwell Campus, Didcot OX11 0DE, United Kingdom}
\author{Urs Staub}
\email{urs.staub@psi.ch}
\affiliation{Center for Photon Science, Forschungsstrasse 111, Paul Scherrer Institute, 5232 Villigen-PSI, Switzerland}

\date{\today}

\begin{abstract}
{Resonant inelastic x-ray scattering (RIXS) has become a prominent technique to study quasiparticle excitations. With advances in polarization analysis capabilities at different facilities, RIXS offers exceptional potential for investigating symmetry-broken quasiparticles like chiral phonons and magnons. At optical wavelengths birefringence can severely affect polarization states in low-symmetry systems. Here we show its importance for soft x-ray resonances. Given the growing interest in Circular Dichroism (CD) in RIXS, it is important to evaluate how birefringence may affect the RIXS spectra of anisotropic systems. We investigate CuO, a well-known anisotropic material, using Cu $L_3$-edge RIXS and detect significant CD in both magnetic and orbital excitations in the collinear antiferromagnetic phase.  We demonstrate that the CD can be modeled by a proper treatment of RIXS scattering amplitudes derived from single-ion calculations with birefringence. Recognizing these effects is crucial for unambiguous identification of subtle dichroic effects induced by symmetry-broken quasiparticles. Furthermore, the combined sensitivity of RIXS and birefringence to local symmetry presents an opportunity to study microscopic changes driven by external perturbations.}     
\end{abstract}



\maketitle
Symmetry constraints of light-matter interactions are fundamental to optical spectroscopic methods. Most materials belong to non-cubic crystal systems and thus exhibit anisotropic responses to light, which is useful for understanding the symmetry based origins of physical properties. At optical wavelengths dichroism (polarization dependent absorption) and birefringence (polarization dependent phase retardation) are two of such commonly utilized anisotropic responses to light~\cite{born_principles_1999}. 

In the x-ray regime, the anisotropic responses are usually negligible since the refractive index is very close to unity, the value for vacuum.  Near the resonant transitions from core shells to valence states in atoms, however, the polarization states of x-rays become highly sensitive to local symmetry. Dichroism spectroscopy with linearly polarized X-ray can be used to investigate orientational order of anisotropic motifs, e.g., X-ray Natural Linear Dichroism, to study element specific orbital orientation along corresponding symmetry axes~\cite{stohr_orientation_1981} and X-ray Magnetic Linear Dichroism to investigate easy spin axis anisotropy~\cite{van_der_laan_magnetic_1999}. Dichroism spectroscopy with circularly polarized x-rays like  X-ray Magnetic Circular Dichroism and X-ray Natural Circular Dichroism can be used to study element-specific magnetic and charge density anisotropy by detecting time-reversal or space-inversion symmetry-breaking, respectively~\cite{schutz_absorption_1987,alagna_x-ray_1998}. Resonant Elastic X-ray Scattering (REXS) of forbidden reflections with circular polarization is yet another powerful method to identify structural anisotropies like crystal chirality or study the handedness of spin-spiral structures~\cite{tanaka_right_2008,ueda_conical_2022,lang_imaging_2004,ortiz_hernandez_magnetoelectric_2024,mulders_circularly_2010}. While these techniques are limited to ground state anisotropies, the development of RIXS has extended the possibilities to study symmetry-broken quasiparticles like phonons and magnons~\cite{ament_resonant_2011,de_groot_resonant_2024}. For instance, long-sought chiral phonons were recently observed in space-inversion symmetry-broken $\alpha$-quartz using CD in RIXS~\cite{ueda_chiral_2023}.  Circular Dichroic-RIXS (CD-RIXS) has been proposed to probe the local orbital angular momentum and associated Berry curvature in inversion symmetry-broken transition metal dichalcogenides~\cite{schuler_probing_2023}, and also to identify different orbital ordered states based on the symmetry of occupied orbitals~\cite{marra_unraveling_2012}. In principle, the presence of the recently proposed chiral magnons in time-reversal symmetry broken altermagnets~\cite{smejkal_chiral_2023} can also be tested using CD-RIXS.

The wavelength dependent complex linear absorption coefficient of a material can be written as $\mu=\mu'+i\mu''$. The polarization dependence in $\mu'$ ($\mu''$) represents dichroism  (birefringence)~\cite{collins_x-ray_2013,lovesey_effects_2014,lovesey_x-ray_2001}, and can be related to the complex refractive index $\tilde{n}=1-[\lambda/4\pi] \mu''+i[\lambda/4\pi] \mu'$ in this notation. Typically, $Re(\tilde{n})\approx1$ at these wavelengths, and therefore birefringence effects are regularly neglected in X-ray studies. However, near X-ray resonances, anomalous dispersion may lead to significant enhancement in birefringence~\cite{palmer_x-ray_2011,collins_x-ray_2013,joly_birefringence_2012,mertins_x-ray_2004,petcov_x-ray_1990,joly_chirality_2014,lovesey_x-ray_2001,palmer_x-ray_2014}. Birefringence has been utilized to develop molecular anisotropy imaging techniques~\cite{palmer_x-ray_2014}, proposed for making non-diffractive x-ray phase retarders~\cite{palmer_x-ray_2011} and actively searched as a signature of translational symmetry breaking of vacuum in presence of strong electromagnetic fields from the propositions of non-linear quantum electrodynamics~\cite{karbstein_probing_2016}.

While x-ray birefringence can be utilized for its symmetry dependence, it may  also obscure effects from important microscopic symmetry-breaking phenomena, leading to conflicting interpretations. A notable example is from the studies of CuO. Initially studied as a model material to understand the physical properties of cuprate superconductors from the viewpoint of Cu-O electronic structure, it was found to exhibit multiferroicity due to space-inversion symmetry breaking induced by spin-spiral  ordering within the temperature range of 213-230 K~\cite{jacobsen_spin_2018,kimura_cupric_2008}. Many of its properties are easily controllable: the multiferroic phase can be extended to room temperature under pressure~\cite{terada_room-temperature_2022}, magnetic order can be controlled optically ~\cite{johnson_femtosecond_2012}, and chiral magnetic domains and optical activity can be tuned by applying an electric field~\cite{masuda_electric_2021,babkevich_electric_2012}. Crystallographically, CuO has been described in both centrosymmetric (\textit{C}2/\textit{c}) and non-centrosymmetric (\textit{Cc}) monoclinic space groups ~\cite{asbrink_cuo:_1991}. Below 213~K, CuO adopts a collinear antiferromagnetic ordered (AF1) phase. In this phase, the local space and time-inversion symmetries can be broken, making it a candidate for possessing atomic magnetoelectric multipoles as tested recently with neutron scattering~\cite{urru_neutron_2023}. A strong CD was observed in REXS measurements at the antiferromagnetic ordering vector $\mathbf{q}_{\mathrm{AFM}}$ (0.5,  0, -0.5) for the globally time-reversal symmetry invariant AF1 phase, even though CD should be zero in the electric-dipole approximation~\cite{scagnoli_observation_2011,misawa_magnetic_2022}. This observation was initially attributed to electric-dipole magnetic-dipole scattering from the symmetry allowed toroidal moments, one of the atomic magnetoelectric multipoles~\cite{scagnoli_observation_2011}. However, the claim was later challenged due to potential contributions from birefringence~\cite{joly_birefringence_2012,lovesey_effects_2014,collins_x-ray_2013}, leaving the origin of CD in REXS unresolved.

While treatment of x-ray birefringence has been reported for X-ray Absorption Spectroscopy (XAS), x-ray reflectivity and REXS~\cite{palmer_x-ray_2011,collins_x-ray_2013,joly_birefringence_2012,mertins_x-ray_2004,petcov_x-ray_1990,joly_chirality_2014,lovesey_x-ray_2001,palmer_x-ray_2014}, its effects have never been observed in or considered for inelastic x-ray spectroscopy. Given the growing interest in CD-RIXS, it is crucial to determine whether and how birefringence affects the polarization dependence of RIXS spectra in anisotropic systems. In this work, we present the incident polarization dependent RIXS spectra at the Cu $L_3$-edge of CuO. We observe significant CD for both magnetic and orbital excitations in the AF1 phase. We demonstrate that the observed CD in RIXS can be modeled by considering x-ray birefringence resulting in excitation dependent CD, which arises from the anisotropy-induced change in x-ray polarization when x-rays propagate through the material. Consequently, careful consideration must be given to the presence of birefringence and its impact on the quantitative estimation of polarization dependent RIXS spectra of low-symmetry systems.

\begin{figure}
	\centering
	\includegraphics[width=1\linewidth]{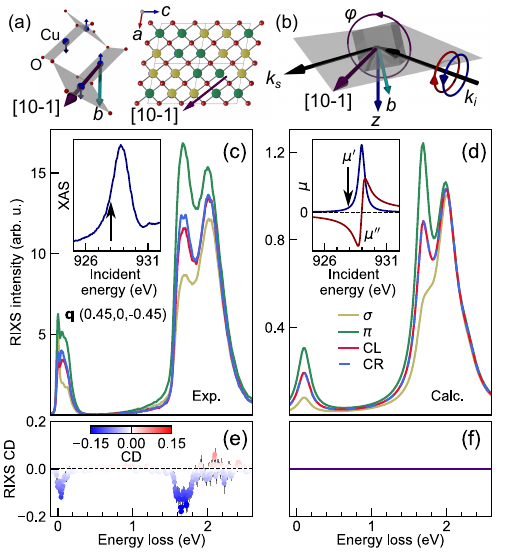}
	\caption{(a) CuO$_4$ plaquettes with anti-parallel spins in the AF1 phase of CuO on the left. On the right, the extended Cu spin arrangement is shown projecting on to the $ac$ plane. The green and yellow spheres represent Cu moments aligned parallel and antiparallel to the $b$ axis. (b) RIXS geometry at I21 beamline, Diamond Light Source, UK~\cite{zhou_i21:_2022}. The $\phi$ motion  rotates the \textit{b}-axis around the [1,0,-1] direction.  (c) RIXS spectra collected on CuO with $\sigma$, $\pi$, CL and CR incident polarizations [color legends are given in panel (d)]. The RIXS intensities are in arbitrary units (arb. u.). Inset shows the Cu $L_3$-edge XAS from CuO and an arrow at $E_{\rm in}=928$~eV. (d) RIXS spectra calculated using the CuO$_4$ plaquettes for the experimental conditions in panel (c). The inset shows the calculated $\mu'$ and $\mu''$  at the Cu $L_3$-edge. (e) CD of excitations observed in RIXS of CuO (see text). The color bar represents deviation from zero CD and the same scale has been used in all the figures. The error bars are standard errors estimated from six independent spectra  for each polarization. (f) Absence of CD in the calculated RIXS spectra. Elastic contributions are not shown for the spectra in panel (d) and the following figures.}
	\label{fig:fig1}
\end{figure}

In the AF1 phase of CuO, the antiparallel spins are aligned to the crystallographic \textit{b}-axis with the ordering vector along [1, 0, -1] direction [see Fig.~\ref{fig:fig1}(a)].
For the RIXS measurements, a CuO single crystal~\cite{prabhakaran_single_2003} was oriented with the  [1, 0, -1] direction along the scattering vector [see Fig.~\ref{fig:fig1}(b) and supplementary materials (SM)~\cite{Supplementary}]. The azimuthal angle ($\phi$) is defined as the angle between the \textit{b}-axis and the laboratory $z$-direction. The temperature was set to 20~K in the AF1 phase. Inset to Fig.~\ref{fig:fig1}(c) shows the XAS with incident polarization normal to the scattering plane ($\epsilon_\perp \equiv \sigma$ polarization) across the Cu $L_3$-edge. RIXS spectra were acquired with incident polarizations $\sigma$ and parallel to the scattering plane ($\epsilon_\parallel \equiv \pi$ polarization) as well as with Circular Left (CL) and Circular Right (CR) polarizations. No polarization analysis of the scattered x-rays was performed. 

The orbital excitations in CuO originate from transitions within the Cu $3d$ orbitals, which have been studied in the past using RIXS, non-resonant inelastic x-ray scattering and quantum chemistry calculations~\cite{wu_effective_2013,ghiringhelli_crystal_2009,huang_ab_2011}. Similar to these works, we have approximated the nearly square CuO$_4$ plaquettes of the highly distorted Cu-O octahedra to have $D_{4h}$ symmetry to simulate the RIXS spectra from CuO using the single-ion multiplet code EDRIXS~\cite{wang_edrixs:_2019} (see SM~\cite{Supplementary}). The magnetic excitations in CuO comprise both well-defined three-dimensional magnons and one-dimensional spinon continuum~\cite{jacobsen_spin_2018}. In our calculations, the magnetic excitations are the local spin-flip amplitudes that scale the intrinsic dynamic spin susceptibility in the RIXS process~\cite{robarts_dynamical_2021}.

In Fig.~\ref{fig:fig1}(c), RIXS  spectra obtained with different incident polarizations at  $\mathbf{q}=$ (0.45,  0, -0.45) with incident energy $E_{\rm{in}}=928$~eV, $T=20$~K and $\phi=60^{\circ}$ are shown. Strong linear dichroism typical of cuprates, is experimentally observed for the magnetic (0.1~eV) and the orbital (1.4-2.4~eV) excitations [see Fig.~\ref{fig:fig1}(c)], which is also replicated in the simulated RIXS spectra shown in Fig.~\ref{fig:fig1}(d). RIXS CD calculated as \rixscdmean, where $I_{\rm{diff}}=I_{\rm{CL}}-I_{\rm{CR}}$ and $\bar{I}_{\rm{sum}}$ is the mean of ($I_{\rm{CL}}+I_{\rm{CR}}$) over the energy loss ($\omega$) range is plotted in Fig.~\ref{fig:fig1}(e). Surprisingly, both the magnetic and the orbital excitations exhibit significant circular contrast. However, the simulated RIXS spectra for CL and CR polarizations are identical [see Fig.~\ref{fig:fig1}(d, f)], as expected from electric-dipole scattering in the AF1 phase. For REXS at $\mathbf{q}_{\mathrm{AFM}}$ contributions from higher-order scattering due to toroidal moments in CuO was estimated to be four orders of magnitude smaller than that of the electric-dipole contributions~\cite{joly_birefringence_2012,di_matteo_orbital_2012}, though absolute estimations of higher-order transitions are not expected to be accurate. Given that CuO is a monoclinic (biaxial) system, we therefore investigate whether the observed  RIXS CD is due to birefringence.
\begin{figure}
	\centering
	\includegraphics[width=1\linewidth]{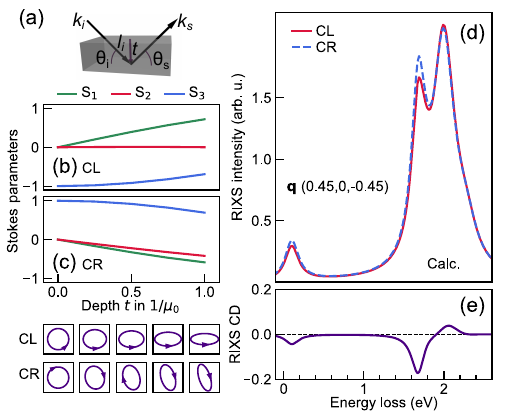}
	\caption{(a) Sketch of x-rays scattered from a region at depth \textit{t} in the sample. Calculated depth dependence of Stokes parameters $S_1$, $S_2$ and $S_3$ for $E_{\rm in}=928$~eV at $\phi=60^{\circ}$ and having initial polarizations (b) CL and (c) CR. The bottom panels display the corresponding polarization ellipses at selected depths to visualize the deviation from the circular polarization. (d) RIXS spectra calculated for CL and CR polarizations using Eq.~\ref{eq:IRIXS} including birefringence contributions. (e) CD in the calculated RIXS spectra.}
	\label{fig:fig2}
\end{figure}
\begin{figure*}[btp]
	\centering
	\includegraphics[width=1\linewidth]{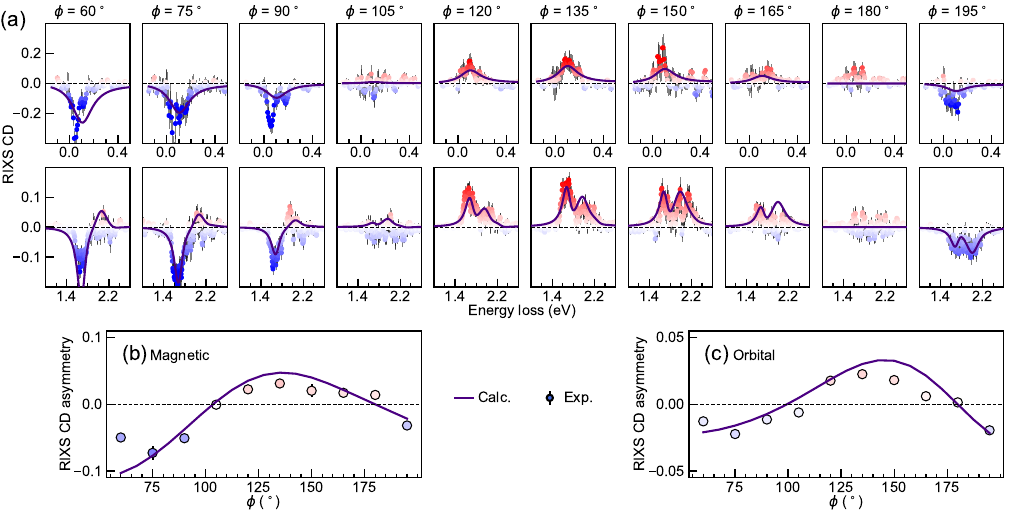}
	\caption{(a) Azimuthal dependence of RIXS CD for ${\bf q}=$ (0.45, 0, -0.45) with $E_{\rm in}=928$~eV. The top (bottom) panels show separate energy ranges of the magnetic (orbital) excitations. Experimental data are denoted by markers. Calculated values using Eq.~\ref{eq:IRIXS} are shown as solid curves. (b) Magnetic and (c) orbital $\phi$ dependent RIXS CD asymmetry obtained by choosing energy ranges of -0.15 to 0.5 and 0.5 to 2.75~eV, respectively (see text).}
	\label{fig:fig3}
\end{figure*}
\begin{figure*}[btp]
	\centering
	\includegraphics[width=1\linewidth]{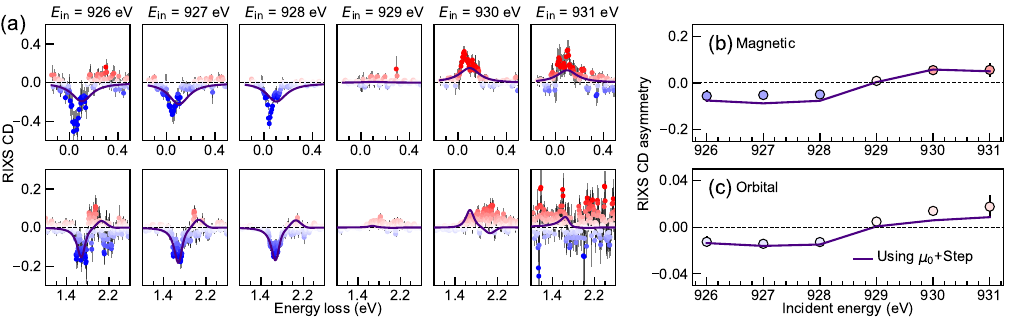}
	\caption{(a) Incident energy dependence of RIXS CD for ${\bf q}=$ (0.45, 0, -0.45) at $\phi=60^{\circ}$. The top (bottom) panels show separate energy ranges of the magnetic (orbital) excitations. Experimental data are denoted by markers. (b) Magnetic and (c) orbital $E_{\rm in}$ dependent RIXS CD asymmetry obtained by choosing energy ranges of -0.15 to 0.5 and 0.5 to 2.75~eV, respectively. Calculated values using Eq.~\ref{eq:IRIXS} are shown as solid curves corresponding to a penetration depth of $1/(\mu_0+\mathrm{step})$.}
	\label{fig:fig4}
\end{figure*}
In RIXS simulations, during x-ray propagation through a material, the only effect considered in previous works is the effect of absorption while the polarization state is kept intact. To model the birefringence or the depth dependent polarization change, we use the transfer matrix formalism developed by Lovesey and Collins for x-rays based on Jones calculus~\cite{jones_new_1948,lovesey_photon_1993,lovesey_x-ray_2001}, and adapt the matrices to calculate RIXS single-ion atomic scattering factors. The polarization matrix in the $\sigma$-$\pi$ basis is written as~\cite{lovesey_photon_1993}:
\begin{equation}
	P=\frac{1}{2}\begin{pmatrix}
		1+S_{1} & S_{2}-i S_{3}\\ 
		S_{2}+i S_{3} & 1-S_{1}
	\end{pmatrix}~
	\label{eq:Stokes}
\end{equation}
where, the Stokes parameters $S_1$, $S_2$ and $S_3$ define the polarization state of light ($S_1=1\Rightarrow\sigma$, $S_1=-1\Rightarrow\pi$, $S_2=\pm1\Rightarrow\pm45^{\circ}$ to $\sigma$ and $\pi$, $S_3=1\Rightarrow \mathrm{CR}$, $S_3=-1\Rightarrow \mathrm{CL}$). After propagating a distance $l_{\rm i} = t/\sin \theta_{\rm i} $ [$\rm i \equiv$ incident, see Fig.~\ref{fig:fig2}(a)], the polarization state $P_{l_{\rm i}}\propto T_{\rm i}PT_{\rm i}^\dag$. The transmittance matrix $T_{\rm i}$ ($T_{\rm i}^\dag$ is its Hermitian conjugate) for the incident x-rays is written out of the $\sigma$-$\pi$ tensorial components of $\mu$~\cite{lovesey_x-ray_2001,lovesey_photon_1993,joly_birefringence_2012}:
\begin{equation}
	T_{\rm i}=e^{-\frac{1}{4}[\mu_{\rm i\pss}+\mu_{\rm i\ppp}]l_{\rm i}}\begin{pmatrix}
		T_{\rm i\pss} & T_{\rm i\pps}\\ 
		T_{\rm i\psp} & T_{\rm i\ppp}
	\end{pmatrix} {\mathrm {where,}}
	\label{eq:T_mat}
\end{equation}
\begin{gather*}
	T_{\rm i\pss(\ppp)} =\cosh \tau_{\rm i} l_{\rm i}+\frac{\mu_{\rm i\ppp(\pss)}-\mu_{\rm i\pss(\ppp)}}{4 \tau_{\rm i}} \sinh \tau_{\rm i} l_{\rm i}~,\\
	T_{\rm i\psp(\pps)} =-\frac{\mu_{\rm i\psp(\pps)}}{2 \tau_{\rm i}} \sinh \tau_{\rm i} l_{\rm i}~,~\mathrm{and}\\
	\tau_{\rm i}=\frac{1}{4} \sqrt{[\mu_{\rm i\pss}-\mu_{\rm i\ppp}]^{2}+4 \mu_{\rm i\psp} \mu_{\rm i\pps}} \nonumber ~~.\\
\end{gather*}
The Stokes parameters are evaluated for CL and CR incident x-ray polarization having $E_{\rm{in}}=928$~eV for $\phi=60^{\circ}$, at different depths $t$ in units of $1/\mu_0$, where $\mu_0$ is the energy dependent isotropic absorption coefficient. The Stokes parameters and the resulting polarization ellipses are shown in Fig.~\ref{fig:fig2}(b) and (c). The deviation from the initial polarization state is evident as the x-ray propagates through the material, which then must be considered for calculating depth dependent RIXS scattering amplitudes. To account for the birefringence induced change in polarization~\cite{joly_birefringence_2012}, we calculate the final RIXS intensity (${\rm s} \equiv$ scattered) as:
\begin{multline}
	I_{\rm RIXS}({\bf k_{i}}, {\bf k_{s}}, {\boldsymbol \epsilon_{\bf i}}, {\boldsymbol \epsilon_{\bf s}},E_{\rm in}, \omega) \propto  \\
	\sum\limits_{f} \int_{0}^{\frac{1}{\mu_0}}{\rm Tr}(T_{\rm s} R_{fg} T_{\rm i} P T_{\rm i}^\dag R_{fg}^\dag T_{\rm s}^\dag)dt   \\ \times \delta (E_f-E_g-\omega)~
	\label{eq:IRIXS}
\end{multline}
where, the RIXS scattering amplitude  in the $\sigma$-$\pi$  basis, 
\begin{equation}
	R_{fg}=\begin{pmatrix}
		R_{fg\pss} & R_{fg\pps}\\ 
		R_{fg\psp} & R_{fg\ppp}
	\end{pmatrix}~.
	\label{eq:R_mat}
\end{equation}
The polarization dependent components in Eq.~\ref{eq:R_mat} are calculated for the plaquettes as:
\begin{eqnarray}
	&& R_{fg {\bf k_{i}} {\bf k_{s}} \boldsymbol \epsilon_{\bf i} \boldsymbol \epsilon _{\bf s}} =  \sum_n
	\frac{\langle f | {D_f}^\dag | n \rangle  \langle n | {D_g} | g \rangle }{E_g-E_n+ E_{\rm in}+i\Gamma_n} ~
	\label{eq:K-H}
\end{eqnarray}
where, \textit{g}, \textit{n} and \textit{f}, and $E_g$, $E_n$ and $E_f$ are the initial, intermediate and final states and their energies in the RIXS process, respectively.  $D_g$ and $D_f^\dag$ are the x-ray polarization and direction dependent transition operators. While $T_{\rm i}$ [Eq.~\ref{eq:T_mat}] is composed of a single  $E_{\rm in}$, the transmittance matrix after the scattering $T_{\rm s}$  has multiple scattered photon energies $E_{\rm in}-\omega$. For polarization resolved RIXS, an analyzer matrix has to be added in the integrand of Eq.~\ref{eq:IRIXS}. Since in this work we do not decompose the scattered x-ray polarization, usage of  $T_{\rm s}$ may seem unnecessary. However, the inclusion of  $T_{\rm s}$ also accounts for the energy dependent self-absorption of the scattered photons.

Fig.~\ref{fig:fig2}(d) shows the RIXS spectra calculated using Eq.~\ref{eq:IRIXS} for CL and CR polarizations and the same experimental conditions as in Fig.~\ref{fig:fig1}(c).  The resulting RIXS CD is plotted in Fig.~\ref{fig:fig2}(e). Considering the polarization state change induced by birefringence, CD is now observed for both the magnetic and the orbital excitations in excellent agreement with the experiment. To demonstrate the validity of our approach, we present in Fig.~\ref{fig:fig3}(a), the experimentally obtained RIXS CD at different $\phi$s for  ${\bf q}=$ (0.45, 0, -0.45) and $E_{\rm in}=928$~eV. The markers in Fig.~\ref{fig:fig3}(b) and (c) show the experimental $\phi$ dependence of RIXS CD asymmetry $\equiv$ \rixsasmean~for the magnetic and the orbital excitations, respectively. The solid curves in the panels show the RIXS CD calculated using Eq.~\ref{eq:IRIXS}. Experimentally, the CD signal for both types of excitations goes from negative to positive and then negative again in the probed $\phi$ range. The calculated RIXS CD follows the experimental $\phi$ dependence excellently both in sign as well as in magnitude as shown by the line profiles in (a) panels and the summed values in panels (b) and (c).



Further confirmation of the birefringence origin of the RIXS CD is obtained from its $E_{\rm{in}}$ dependence presented in Fig.~\ref{fig:fig4}(a) for ${\bf q}=$ (0.45, 0, -0.45) at $\phi=60^{\circ}$. The RIXS CD changes sign as $E_{\rm in}$ is scanned across the Cu $L_3$-edge. For a $d^9$ system the absorption peak comprises of a resonant excitation to a single multiplet state. It is therefore unlikely for any other symmetry-breaking phenomena (like the development of toroidal moments) to induce such a sign change due to interference among different multiplet states.  Instead, $\mu''$ which is related by the Kramers-Kronig transformation to $\mu'$, undergoes a sign change across the absorption edge [see inset to Fig.~\ref{fig:fig1}(d)]. At the resonance, where $\mu''$ is zero, the birefringence contribution is zero and so is the RIXS CD. The experimentally observed sign change across the Cu L$_3$-resonance therefore appears in our calculations when birefringence is included. Note that the RIXS CD is also observed in the paramagnetic phase at $T=240~K$ [see SM~\cite{Supplementary}], also supporting the birefringence interpretation.

Using the framework developed for birefringence in RIXS with single-ion calculations we were also able to simulate the REXS CD of CuO at 
$\mathbf{q_{\mathrm{AFM}}}$~\cite{scagnoli_observation_2011,misawa_magnetic_2022} like earlier theoretical works~\cite{joly_birefringence_2012,lovesey_effects_2014,collins_x-ray_2013} (see SM~\cite{Supplementary}). However, it should be noted that away from the XAS peak at $E_{\rm in}=929$~eV the calculated  RIXS CD grows rapidly compared to the experiment (see SM~\cite{Supplementary}). This is due to the strongly decaying single-ion absorption that results in a rapid enhancement of the penetration depth (1/$\mu_0$ in Eq.~\ref{eq:IRIXS}). In the real material contribution from oxygen and continuum states represented by a step function over the edge,  lead to a slower variation of $\mu_0$ away from the resonance where the polarization change occurs~\cite{henke_x-ray_1993}. When a step function representing the transitions to these states is considered so that the penetration depth is $1/(\mu_0+\mathrm{step})$, we observe a significant improvement in the energy dependent RIXS CD from Fig.~\ref{fig:fig4} (b) and (c) (see SM for the details~\cite{Supplementary}). This suggests that although our calculations do not contain any free scaling parameter, to exactly determine the magnitude of birefringence-induced RIXS CD, precise knowledge of the energy-dependent penetration depth is crucial. This is unfortunately difficult to obtain experimentally for insulating materials like CuO (electron yield is preferable over fluorescence yield to minimize self-absorption and birefringence effects), or one needs inputs from first-principles electronic structure theories~\cite{joly_birefringence_2012}. Therefore, our calculations do not invalidate a finite presence of other symmetry-breaking contribution to the observed RIXS or REXS CD~\cite{scagnoli_observation_2011,urru_neutron_2023}.  In addition, experimentally sampling multiple domains can lead to scaling of the CD magnitude ~\cite{misawa_magnetic_2022,Supplementary}, further increasing the uncertainty in estimating the contribution from the higher-order scatterings.

To conclude, we demonstrate experimentally the presence of significant CD in the RIXS spectra of CuO at the Cu $L_3$-edge in its collinear antiferromagnetic phase. Our simulations show that the CD can be well-described by birefringence from modelling the RIXS scattering amplitudes derived from single-ion models, as detailed in  Eq.~\ref{eq:IRIXS}. This approach offers a framework for detecting or predicting the impact of birefringence in RIXS experiments. This finding emphasizes the need to account for birefringence when performing quantitative polarization analysis of RIXS spectra (including linear dichroism) from any low-symmetry system.  Experimental conditions should therefore be chosen carefully to mitigate the birefringence effects while searching for other subtle symmetry-breaking phenomena associated with the quasiparticle excitations. Propagating x-rays along the highest symmetry direction can reduce the effect of birefringence for uniaxial systems. While limiting the study to the peak of $L_3$-edge XAS can help eliminate it for $d^9$ systems, and it is important to note that for lower electronic configurations, birefringence may exhibit a more complex energy dependence due to interference among different multiplet states~\cite{ueda_non-chiral_2024,haverkort_symmetry_2010}, requiring more careful considerations. Nonetheless, birefringence in RIXS could serve as a valuable tool in condensed matter physics, allowing the study of quasiparticle dynamics and their response to subtle changes in local symmetry induced by variations in temperature, pressure, or light pulses.

We thank Y. Joly for insightful discussions. We acknowledge Diamond Light Source for providing the beam-time at beamline I21 under Proposal MM26777. A.N. acknowledges the Marie Sk\l odowska-Curie Grant Agreement No. 884104 (PSI-FELLOW-III-3i).  A.N. (in part) and G.S.P. were supported by funding from the Swiss National Science Foundation through Project No. 20021-196964.

\end{document}


\newcommand{\rixscd}{$[I_{\rm{CL}}-I_{\rm{CR}}]/\sum\limits_{\omega}[I_{\rm{CL}}+I_{\rm{CR}}]$}
\newcommand{\rixsas}{$\sum\limits_{\omega}(I_{\rm{CL}}-I_{\rm{CR}})/
\sum\limits_{\omega}(I_{\rm{CL}}+I_{\rm{CR}})$}
\newcommand{\pss}{\sigma\sigma}
\newcommand{\psp}{\sigma\pi}
\newcommand{\pps}{\pi\sigma}
\newcommand{\ppp}{\pi\pi}
\renewcommand{\figurename}{Fig.}
\renewcommand{\thefigure}{S\arabic{figure}}

\title{Circular dichroism in resonant inelastic x-ray scattering from birefringence in CuO}

\author{Abhishek Nag}
\email{abhishek.nag@ph.iitr.ac.in}
\thanks{These authors contributed equally to this work.}
\affiliation{Center for Photon Science, Forschungsstrasse 111, Paul Scherrer Institute, 5232 Villigen-PSI, Switzerland}
\affiliation{Department of Physics, Indian Institute of Technology Roorkee, Uttarakhand 247667, India}
\author{G\'erard Sylvester Perren}
\thanks{These authors contributed equally to this work.}
\affiliation{Center for Photon Science, Forschungsstrasse 111, Paul Scherrer Institute, 5232 Villigen-PSI, Switzerland}
\author{Hiroki Ueda}
\affiliation{Center for Photon Science, Forschungsstrasse 111, Paul Scherrer Institute, 5232 Villigen-PSI, Switzerland}
\author{A. T. Boothroyd}
\affiliation{Department of Physics, University of Oxford, Clarendon Laboratory, Parks Road, Oxford OX1 3PU, UK.}
\author{D. Prabhakaran}
\affiliation{Department of Physics, University of Oxford, Clarendon Laboratory, Parks Road, Oxford OX1 3PU, UK.}
\author{M. Garc\'ia-Fern\'andez}
\affiliation{Diamond Light Source, Harwell Campus, Didcot OX11 0DE, United Kingdom}
\author{S. Agrestini}
\affiliation{Diamond Light Source, Harwell Campus, Didcot OX11 0DE, United Kingdom}
\author{Ke-Jin Zhou}
\affiliation{Diamond Light Source, Harwell Campus, Didcot OX11 0DE, United Kingdom}
\author{Urs Staub}
\email{urs.staub@psi.ch}
\affiliation{Center for Photon Science, Forschungsstrasse 111, Paul Scherrer Institute, 5232 Villigen-PSI, Switzerland}

\date{\today}



\maketitle
\section{Experimental details}

A polished single crystal of CuO grown using floating-zone technique was used for the experiment~\cite{prabhakaran_single_2003}. All measurements were done at I21 beamline, Diamond Light Source, UK~\cite{zhou_i21:_2022}. The sample was cooled down to 20 K for measurements in the AF1 phase. The pressure inside the sample vessel was maintained around $5\times10^{-10}$ mbar. The x-ray absorption (XAS) was collected as fluorescence yield (FY) using a photodiode. High energy-resolution ($\Delta E= 0.035$ eV) RIXS spectra were collected at Cu $L_3$-edge. The zero-energy transfer position and energy resolution were determined from subsequent measurements of elastic peaks from an adjacent carbon tape. 

The vertical (horizontal) x-ray spot size on the sample was around 5 $\mu$m (35 $\mu$m) at normal incidence. The magnetic domain dimensions in CuO crystals have been seen to greater than 100 $\mu$m~\cite{misawa_magnetic_2022}, and thus it is likely that we have sampled majorly a single domain throughout the experiment. 

\begin{figure}[H]
	\centering
	\includegraphics[width=1\linewidth]{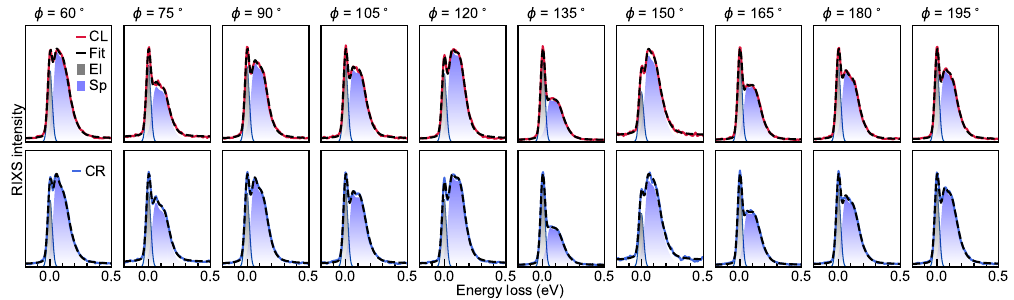}
	\caption{The panels show the low-energy RIXS spectra collected at ${\bf q}=$ (0.45, 0, -0.45), incident energy $E_{\rm in}=928$~eV at different azimuthal angles ($\phi$). The top (bottom) panels are measurements with CL (CR) incident x-ray polarizations. The dashed curves represent the total fit to the features, while the shaded parts represent the elastic and the magnetic contributions.}
	\label{fig:fitting}
\end{figure}

The low-energy RIXS features were fit using a Gaussian line-shape for the elastic peak, and Voigt line-shapes for the magnetic peaks (magnon and spinon contributions~\cite{jacobsen_spin_2018}) [see Fig.~\ref{fig:fitting}]. Using the fitted elastic peak position we could align the CL and CR RIXS spectra to an accuracy of about 0.005~eV. The fitted elastic peak was subtracted from the RIXS spectra when calculating RIXS CD that have been presented in the manuscript. Note that no self-absorption correction was applied to the experimental data. 

\begin{figure}[htbp!]
	\centering
	\includegraphics[width=0.5\linewidth]{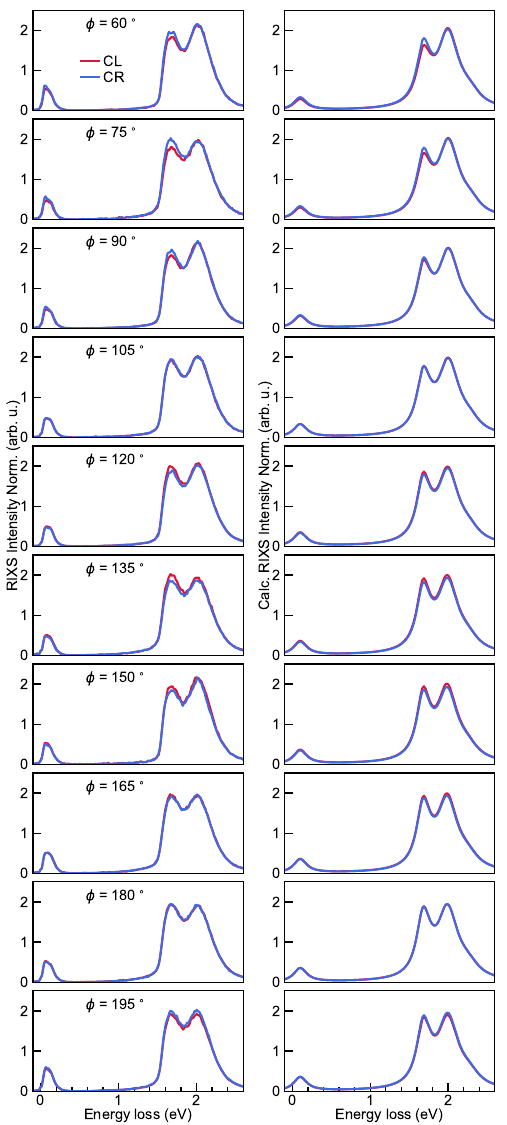}
	\caption{The panels show the RIXS spectra with CL and CR incident x-ray polarizations at ${\bf q}=$ (0.45, 0, -0.45), incident energy $E_{\rm in}=928$~eV for different azimuthal angles ($\phi$). Each spectrum has been normalized by mean of ($I_{\rm{CL}}+I_{\rm{CR}}$) over the energy loss ($\omega$) range. The left (right) panels correspond to experimental (calculated) spectra.}
	\label{fig:azimuth}
\end{figure}

\begin{figure}[htbp!]
	\centering
	\includegraphics[width=0.5\linewidth]{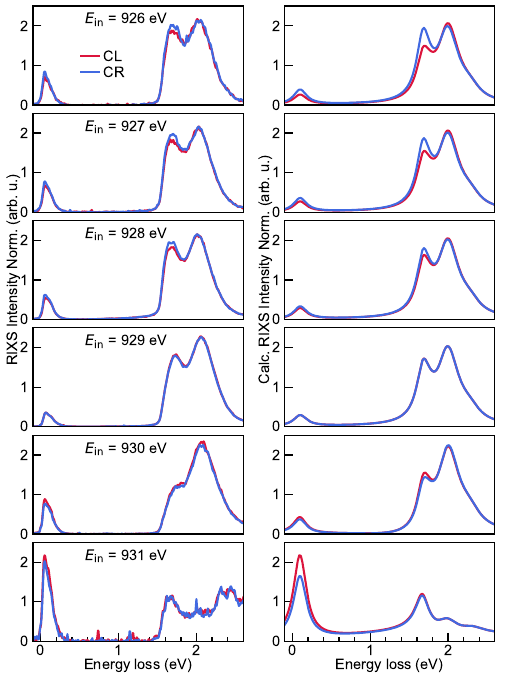}
	\caption{The panels show the RIXS spectra with CL and CR incident x-ray polarizations at ${\bf q}=$ (0.45, 0, -0.45), $\phi=60^{\circ}$ for different incident energies ($E_{\rm in}$). Each spectrum has been normalized by mean of ($I_{\rm{CL}}+I_{\rm{CR}}$) over the energy loss ($\omega$) range. The left (right) panels correspond to experimental (calculated) spectra.}
	\label{fig:energy}
\end{figure}

In the manuscript we have presented the RIXS CD spectra for the azimuthal and energy dependence. The experimental and calculated RIXS spectra corresponding to the azimuthal dependence is shown in Fig.~\ref{fig:azimuth} and the energy dependence is shown in Fig.~\ref{fig:energy}.

\begin{figure}[htbp!]
	\centering
	\includegraphics[width=1\linewidth]{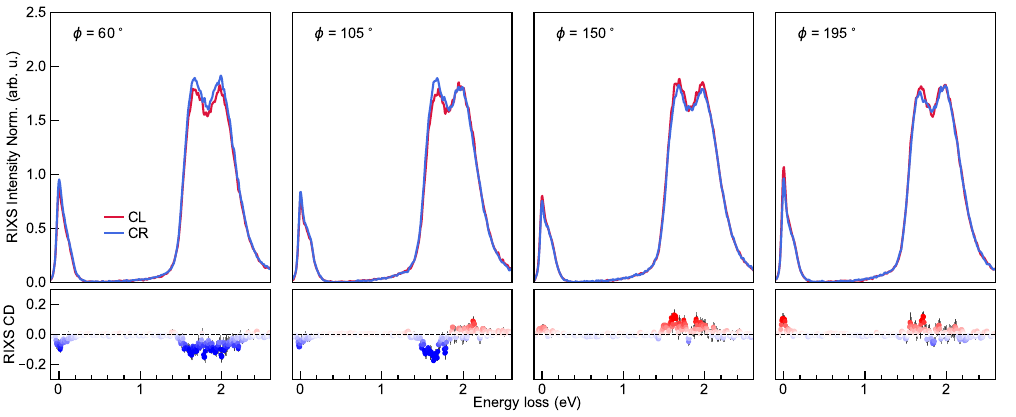}
	\caption{The top panels show the RIXS spectra collected with CL and CR incident x-ray polarizations at ${\bf q}=$ (0.45, 0, -0.45), incident energy $E_{\rm in}=928$~eV and $T=240$~K for different azimuthal angles ($\phi$). Each spectrum has been normalized by mean of ($I_{\rm{CL}}+I_{\rm{CR}}$) over the energy loss ($\omega$) range. The bottom panels show the CD of excitations observed in RIXS of CuO in the paramagnetic phase.}
	\label{fig:Paramagnetic}
\end{figure}

Additional RIXS data were collected at 240~K in the paramagnetic phase at different azimuthal ($\phi$) angles which also show the presence of RIXS CD as shown in Fig.~\ref{fig:Paramagnetic}. Note that the thermal expansion of the sample manipulator between 20~K and 240~K can lead to measurement at a slightly different spot, even though measures were taken to correct for the drift in the sample position.

\section{Theory details}
As mentioned in the manuscript we have approximated the Cu site with a $D_{4h}$ symmetry for the calculations. Similar higher symmetries as compared to $C_1$ in the real material were found to satisfactorily explain earlier non-resonant inelastic x-ray scattering results on CuO~\cite{wu_effective_2013}. The orbital (\textit{dd}) excitations in this case correspond to spin-flip and non-spin-flip excitations to the $d_{xy}$, $d_{yz}$, $d_{xz}$ and the $d_{3z^2-r^2}$ orbitals from the $d_{x^2-y^2}$ orbital. In the past, energies of  around 1.8, 2.1, and 2.2~eV have been attributed to excitations to the $d_{xy}$, $d_{yz}$ and $d_{xz}$, and $d_{3z^2-r^2}$ orbitals respectively, from the investigation of the orbital states in CuO using RIXS, non-resonant inelastic x-ray scattering and quantum chemistry calculations~\cite{wu_effective_2013,ghiringhelli_crystal_2009,huang_ab_2011}. We have considered four plaquettes as shown in Fig.~1(a) of the manuscript to simulate the XAS and RIXS spectra from CuO using the single-ion multiplet code EDRIXS~\cite{wang_edrixs:_2019}. A plaquette dependent directional exchange field was added to the calculations to represent the local spin-flip RIXS amplitude that scales the polarization or $E_{\rm{in}}$ independent dynamic spin susceptibility to give the experimental RIXS spin-wave  intensity~\cite{robarts_dynamical_2021}. 

\begin{figure}[htbp!]
	\centering
	\includegraphics[width=0.5\linewidth]{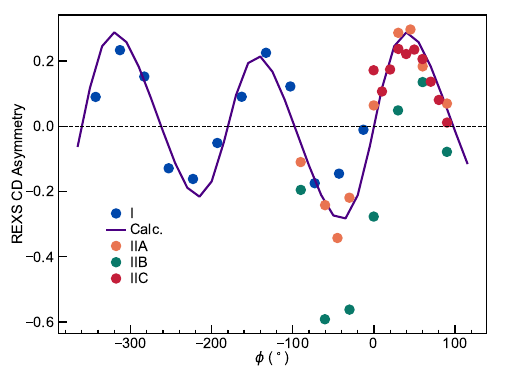}
	\caption{The panel shows experimentally observed REXS CD asymmetry at $\mathbf{q}_{\mathrm{AFM}}=$ (0.5,  0, -0.5) and in the AF1 phase of CuO from Refs.~\cite{scagnoli_observation_2011,misawa_magnetic_2022}. The data I from Ref.~\cite{scagnoli_observation_2011} has been shifted by $90^\circ$ to match the definition of $\phi=0^\circ$ (\textit{b}-axis along the laboratory $z$-direction). The data II are from Ref.~\cite{misawa_magnetic_2022} with different x-ray spot sizes. IIA: Average over unequal amounts of magnetic domains probed with focused beam. IIB: X-ray beam focused on a single domain. IIC: Unfocused x-ray beam over unequal amounts of magnetic domains. The solid line represents the birefringence included REXS CD calculated at 2.7~eV below resonance, using the single-ion formalism presented in the manuscript.}
	\label{fig:rexs}
\end{figure}

For the single-ion calculations, we have used Slater-Condon parameters of $F^2_{dd}=12.854~(13.611)$~eV, $F^4_{dd}=7.980~(8.457)$~eV, $F^2_{pd}=(8.177)$~eV, $G^1_{pd}=(6.169)$~eV, $G^3_{pd}=(3.510)$~eV for the intial (intermediate) states. The $3d$ spin-orbit coupling was taken as 0.102 (0.124)~eV for the intial (intermediate) states. The values of these parameters were scaled to 80\%. The $2p$ spin-orbit coupling was taken as 13.498~eV. The core-hole half-width broadening was taken as 0.28~eV. The crystal field parameters used were $10Dq=1.68$~eV, $Dt=0.146$~eV and $Ds=0.35$~eV. We did not extensively fine-tune the parameters, as they were adequate to demonstrate the effects of polarisation. Note that the origin of RIXS CD does not depend on the exact values of these parameters. For the REXS CD calculations shown in Fig.~\ref{fig:rexs}, we used 16 sites representing the commensurate antiferromagnetic CuO unit cell.

\begin{figure*}[htbp!]
	\centering
	\includegraphics[width=1\linewidth]{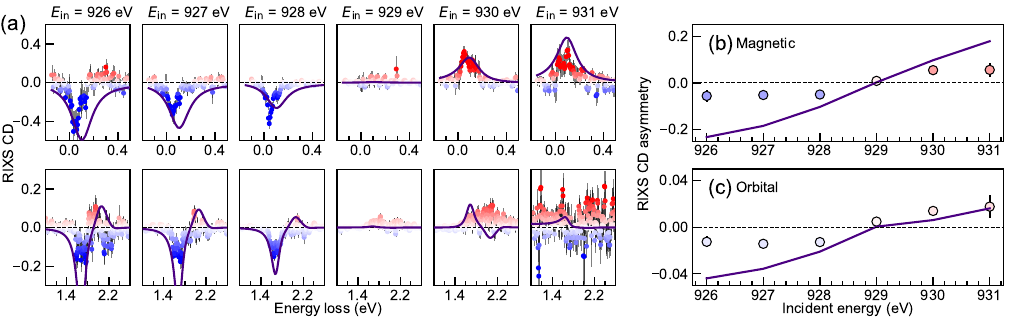}
	\caption{(a) Incident energy dependence of RIXS CD for ${\bf q}=$ (0.45, 0, -0.45) at $\phi=60^{\circ}$. The top (bottom) panels show separate energy ranges of the magnetic (orbital) excitations. Experimental data are denoted by markers. (b) Magnetic and (c) orbital $E_{\rm in}$ dependent RIXS CD asymmetry obtained by choosing energy ranges of -0.15 to 0.5 and 0.5 to 2.75~eV, respectively. Calculated values using Eq. 3 of the manuscript are shown as solid curves corresponding to a penetration depth of $1/(\mu_0)$.}
	\label{fig:SI_fig_mu0_CD}
\end{figure*}

\begin{figure}[htbp!]
	\centering
	\includegraphics[width=0.5\linewidth]{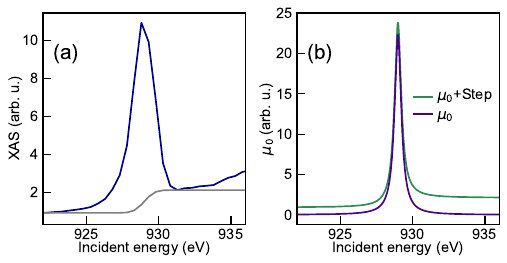}
	\caption{(a) Experimentally obtained XAS on CuO in FY mode and a step function approximating the edge-jump. (b) $\mu_0$ obtained from single-ion calculations with and without the step function shown in (a) for calculating the RIXS CD in Fig.~4 of the manuscript (see text).}
	\label{fig:mu0step_xas}
\end{figure}

In Fig.~\ref{fig:SI_fig_mu0_CD}, it can be seen that away from the XAS peak at $E_{\rm in}=929$~eV, the calculated RIXS CD for a penetration depth of $1/\mu_0$ grows rapidly compared to the experiment. Fig.~\ref{fig:mu0step_xas}(a) shows the experimentally obtained asymmetric XAS in FY mode. To include the effect of background contribution from oxygen and continuum states on RIXS CD, we added a step function to $\mu_0$ [Fig.~\ref{fig:mu0step_xas}(b)] for estimating the total penetration depth of $1/(\mu_0+\mathrm{step})$. We estimated the step height by adjusting the step function before and after the white-line peak in XAS [Fig.~\ref{fig:mu0step_xas}(a)]. We then scaled the $\mu_0$ calculated in the same configuration as the experiment to the XAS and added the estimated step function to it [Fig.~\ref{fig:mu0step_xas}(b)]. The RIXS CD shown in Fig.~4 of the manuscript was then calculated till penetration depths of $1/(\mu_0+\mathrm{step})$ using  Eq.~3 of the manuscript. It is clear that the small XAS background leads to a reasonably good match with the experiment. However, this also shows that away from resonance, a small change in penetration depth can lead to deviation from the single-ion estimated magnitude of the birefringence induced CD.



%